\documentclass{iaus}
\usepackage{graphicx}

\title[The structure of the accretion disk in NGC~4258] %% give here short title %%
{ The structure of the accretion disk in NGC~4258 derived from observations of its water vapor masers }

\author[Moran, Humphreys, Greenhill, Reid \& Argon]   %% give here short author list %%
{James M. Moran$^1$%
  \thanks{Present address: 60 Garden Street, MS-42, Cambridge, MA  02138 USA},
 Elizabeth Humphreys, Lincoln J. Greenhill, \break Mark Reid \and Alice Argon}

\affiliation{$^1$Harvard-Smithsonian Center for Astrophysics \break email: jmoran@cfa.harvard.edu}

\pubyear{2007}
\volume{242}  %% insert here IAU Symposium No.
\pagerange{xxx--xxx}
\date{?? and in revised form ??}
\jname{Astrophysical Masers and their Environments}
\editors{Jessica Chapman \& Willem Baan, eds.}
\begin{document}

\maketitle

\begin{abstract}
A wealth of new information about the structure of the maser disk in NGC~4258 has been obtained from a series of 18 VLBA observations spanning three years, as well as from 32 additional epochs of spectral monitoring data from 1994 to the present, acquired with the VLA, Effelsberg, and GBT.  The warp of the disk has been defined precisely.  The thickness of the maser disk has been measured to be 12 microarcseconds (FWHM), which is slightly smaller than previously quoted upper limits.  Under the assumption that the masers trace the true vertical distribution of material in the disk, from the condition of hydrostatic equilibrium the sound speed is 1.5 km~s$^{-1}$, corresponding to a thermal temperature of 600K.  The accelerations of the high velocity maser components have been accurately measured for many features on both the blue and red side of the spectrum.  The azimuthal offsets of these masers from the midline (the line through the disk in the plane of the sky) and derived projected offsets from the midline based on the warp model correspond well with the measured offsets.  This result suggests that the masers are well described as discrete clumps of masing gas, which accurately trace the Keplerian motion of the disk.  However, we have continued to search for evidence of apparent motions caused by ``phase effects.''  This work provides the foundation for refining the estimate of the distance to NGC~4258 through measurements of feature acceleration and proper motion.  The refined estimate of this distance is expected to be announced in the near future. 

\keywords{NGC~4258, galaxies: active, accretion disks, masers, galaxies: distances and redshifts, black hole physics}

\end{abstract}

\firstsection % if your document starts with a section,
\section{Introduction}

A group based largely at the CfA has been working for the 
past decade to better understand the properties of the 
remarkable accretion disk in NGC~4258 that is traced by its water 
vapor maser emission (for a general review of its properties, see Moran, Greenhill \& Herrnstein, 1999). In addition to the recent work reported here, important results have been obtained by a series of research
projects by Harvard graduate students: Ann Bragg (acceleration analysis; Bragg et al., 2000), James Herrnstein (warp analysis; Herrnstein 1997, Herrnstein et al. 1997, 1998, 2005 and precise distance estimate; Herrnstein et al. 1999), Maryam Modjaz (magnetic field upper limit; Modjaz et al. 2005), and Adam Trotter (warp analysis; Herrnstein et al., 2005). There are two principal foci of the current work: (1) to improve the  understanding of the physics of the accretion disk, and (2) to model precisely the dynamics to allow for the determination of an accurate distance to the galaxy, and to guide the modeling of other galaxies that have  more limited maser data. Here, we describe some of the refinements to the basic model. The distance determined from the accelerations and proper motions of the maser features under the assumption of circular orbits is 7.2 $\pm$ 0.5 Mpc (Herrnstein et al., 1999; compare cepheid distance of 7.5 $\pm$0.25 Mpc for an LMC distance of 50 kpc by Macri et al., 2007). At this (maser distance) 1 mas = 0.035 pc = 1.1 $\times$ 10$^{17}$ cm. For convenience, most radii are  given in angular units in this paper. 

The principal dataset is a series of 18 VLB experiments (VLBA only, plus VLBA augmented by large dishes) undertaken over a period of three years from March 1997 to August 2000. These data have been systematically analyzed, described in the literature (Argon et al., 2007) and the tabulated results made available in the on-line version of the paper. In all, a total of about 14,000 features (defined here as velocity channels) were imaged and their positions, intensities, line widths and accelerations reported (3,967 features in the redshifted group of features, 267 in the blue group, and 10,018 in the systemic group). A critical contribution of this dataset is the measurement of many new features in the blueshifted group, which are much weaker and underrepresented compared to the red group. (A possible explanation for persistent weakness of the blue masers with respect to the redshifted masers is that, because of the warp and orientation of the disk, line of sight to the blue masers passes through the foreground disk and may suffer substantial attenuation in an ionized layer of gas.) The measurement of these masers will provide a much more accurate determination of parameters (e.g., center of symmetry) (Humphreys et al., 2007a).  In addition, we also analyzed spectra from the VLA, GBT and Effelsberg to obtain feature accelerations.
\vspace{-8pt}

\begin{figure}[!hb]
\begin{center}
\includegraphics[height=3.7in,width=2.8in]{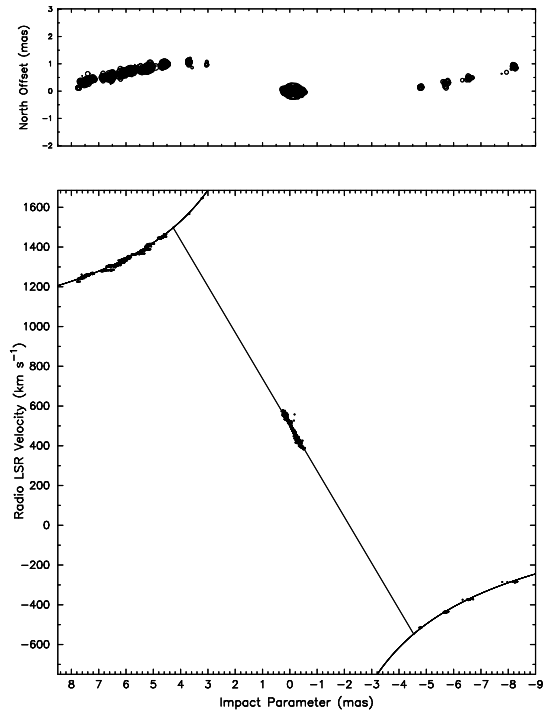}
\vspace{10pt}
  \caption{ (top) Image of the masers in NGC~4258 (1 mas = 0.035 pc). 
(bottom) Position-velocity diagram of the masers. Note that
the masers at 1556 and 1647 km~s$^{-1}$ lie inside the point where the line through the systemic features
intersects the Keplerian curve, which indicates  that they lie at smaller  radii than the  
systemic features. From Argon et al. (2007).}
\end{center}
\end{figure}
\vspace{-8pt}
\section{The warp of the disk}\label{sec:warp} 

Figure 1 shows the image of the masers and the position-velocity (PV) diagram computed along the major axis of the maser distribution. Since the accelerations of the high velocity features are close to zero, these masers lie along the so-called ``midline''  (the line through the disk from the black hole which is in the 
plane perpendicular to the line of sight).  The anti-symmetric distribution of masers is readily understood as  a ``position-angle'' warp. The high velocity portions of the PV diagram follow Kepler's law to an astonishing accuracy of about 1 percent. However, the velocities show a slight flattening, which can be
thought of as   systematic deviations from the Keplerian prescription 
by about 8 km~s$^{-1}$. The most reasonable explanation of this effect is that the disk has a component of ``inclination-angle warp.'' The warp is fully characterized by these two components (i.e., $\alpha$ is the position angle of the disk measured West from North, $i$ is the inclination angle measured from North),  and has been parameterized in the form (in degrees): 

\begin{eqnarray}
\alpha & = & 66+5.0r-0.22r^2\\
  \phi & = & 107-2.29r
\end{eqnarray}\vspace{10pt}

\noindent where $r$ is the radius in mas, by Herrnstein et al. (2005) based on pre-1997 data. Hence, the disk is defined by y = $r$~cos $\alpha$ and y = $r~$sin$~i$ in the plane of the sky and along the line of sight, respectively. This  parameterization reveals three important characteristics of the disk as we view it. Firstly, there is a point 7.4 mas from the center of the disk where the inclination crosses 90$^{\circ}$, i.e., the disk is seen  edge on at this radius. Inside this radius, the front of the disk tips down; outside, it tips up. Hence, the extreme outer portions of the high velocity emission are seen towards the underside of the disk.  Secondly, the warp creates a shallow ``bowl'' on the front side of the maser that has a minimum point  at a radius of 3.9 mas along the line of sight from the black hole. Finally, along the line of sight, the inclination angle becomes 90$^{\circ}$ at a radius of 7.4 mas. The gas at this location may be responsible for the thermal absorption of the x-ray emission (Fruscione et al., 2005).

\begin{figure}
\includegraphics[height=5.18in, width=4.00in, angle=-90]{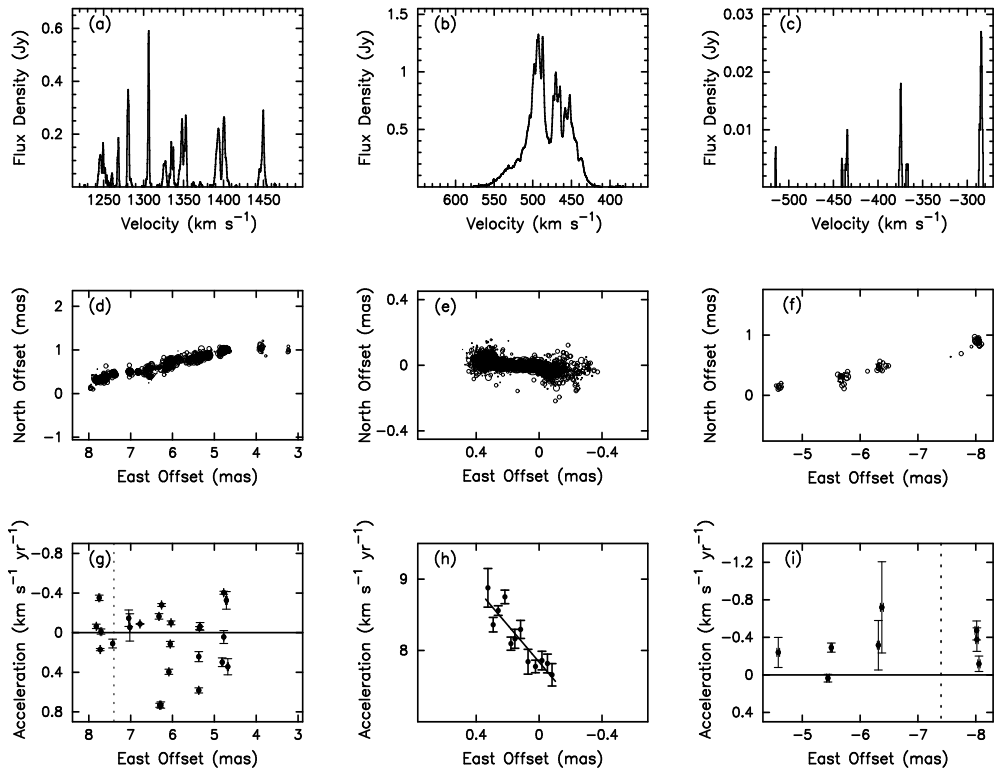}
\vspace{10pt}
\caption{ Spectra (top row), positions (middle) and accelerations (bottom) of the redshifted (left column), systemic (middle), and blueshifted masers (right). Note that the acceleration data for the 
systemic masers has been binned.  The offset distance is measured from the position of the black hole. Note changes in acceleration scales for different velocity groups. The apparent periodicities in velocities (panels a and c) and positions (panels d and f) of red and blueshifted features may indicate spiral structure as suggested by Maoz (1995). Adapted from Argon et al. (2007) and Humphreys et al. (2007b).}
\end{figure}

\section{Acceleration of the systemic features} 

The positions, velocities and accelerations of the systemic features are shown in Figure 2. The spectrum of the systemic maser group is very crowded with features, which makes it difficult to track them individually because of changes in lineshape and intensity and because of limited temporal sampling. 
 It has long been known that the systemic features drift at about 8 km~s$^{-1}$ yr$^{-1}$ (see, e.g., Haschick \& Baan 1990 and Greenhill et al., 1995), traversing the observed masing region  in about 10 years (from this, the beam angle of the masers has been deduced to be about 8$^{circ}$).  Humphreys et al. (2007b) have found  a global solution for a set of Gaussian components, each of which is assumed to be drifting at a constant velocity.  They show that there is an approximate linear change in the acceleration with position in the systemic features (see Figure 2-h). A simple explanation of this tendency may be that the masing region has a fixed dependence of the radius of the masing region as a function of position. The line-of-sight component of the acceleration is given by 
 
 \begin{equation}
  a= \frac{GM}{r^2} \cos \theta~,
  \end{equation}
 where $G$ is the gravitational constant, $M$ is the black hole mass, $r$ is the radius, and $\theta$ is the azimuth angle measured from the line of sight eastward.  Hence, the change in acceleration from 7.7 to 8.8 km~s$^{-1}$ yr$^{-1}$ may be due to a change in radius from 4.0 to 4.2 mas (the azimuth angles are close to zero). Note that the measurements of acceleration from earlier epochs seem to show this same dependency, which suggests that this pattern does not move with the disk rotation.  However, there is a more intriguing explanation for this dependency. Perhaps the accretion disk contains  a spiral density wave, with a pattern speed less than the rotation speed, which, because of its overdensity provides an extra component of acceleration. A model with a spiral arm mass of about $10^4$ M$_{\odot}$ (about 15 percent of the maximum allowable disk mass), could account for the trend in acceleration with position (Humphreys et al., 2007b). If this model is correct,  changes in the acceleration might allow the pattern speed to be measured. Finally, it is conceivable, though theoretically improbable (see, e.g., Statler et al., 2001), that the maser orbits are confocal ellipses. Tight limits on any eccentricity in the orbits have been placed by Humphreys et al. (2007b).

\begin{figure}[!ht]
\begin{center}
\includegraphics[height=3.7in,width=3.7in,angle=-90]{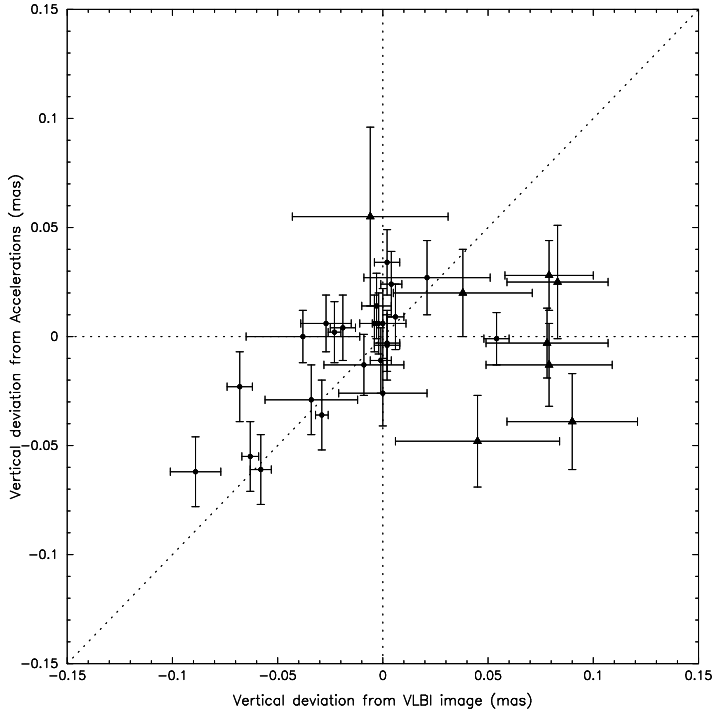}
\caption{Comparison of expected projected vertical positions based on the azimuth angles calculated from accelerations and projection 
based on the warp versus the observed projected vertical positions.  >From Humphreys et al. (2007b).}
\end{center}
\end{figure}

\section{Acceleration of the high velocity features}

The accelerations of the high velocity features (see Figure 2) are about a factor of twenty smaller than those of the systemic masers. The origin of these accelerations is probably the result of a purely geometric effect. That is, masers lying on the midline will have exactly zero acceleration, while those off the midline  will have accelerations (see Eq. 3.1) of $a \sim GM(90-\theta)/r^2$ . With this interpretation, the spread of about $\pm$ 1.0 km~s$^{-1}$ yr$^{-1}$ leads to a range of inferred azimuth angles of $\pm$ 10 degrees. If these azimuth angles are correct and the parameters of the warp model are accurate, then there should be a strong correlation between the offset of the masers from the local midplane as calculated from the azimuth angle and the warp, and the measured position offsets. Figure 3 shows that these position estimates are highly correlated. The correlation is not perfect and is weakest for the blue features. However, the warp model parameters are largely determined by the redshifted features. Furthermore, the finite thickness of the disk of about 12 $\mu$as (see section 5 below) will degrade the correlation. We consider this correlation to be strong evidence that the masers are accurately modeled as ballistic clumps and that the warp, as parameterized in Eq. 2.1, has been reasonably well modelled. 

A model of a spiral density wave moving through the accretion disk could create apparent accelerations, i.e., the radial motion of the spiral density wave causes the excitation point in the masing gas to move outward in time causing a decrease in velocity on the redshifted side of the disk as a function of time and a blue positive drift in velocity on the blueshifted side.  The magnitude of the effect is $a_{app} = \pm 0.05(\theta_p/2.5^{\circ})$ km~s$^{-1}$yr$^{-1}$, where $\theta_p$ is the pitch angle, the positive sign refers to the blueshifted masers and the negative sign to the redshifted ones (Maoz \& McKee 1998). This effect may be present, but it is much smaller than, and masked by, the geometric effect.  Note, however, that the mean value of the accelerations seems to be slightly positive on the red side and significantly negative on the blue side, the opposite of what is expected for a trailing spiral.  Our conclusion is contrary to that of Yamauchi et al. (2005), but their analysis is based on only a single blueshifted acceleration.

\begin{figure}[!hb]
\begin{center}
\includegraphics[height=3.7in, width=2.8in, angle=270]{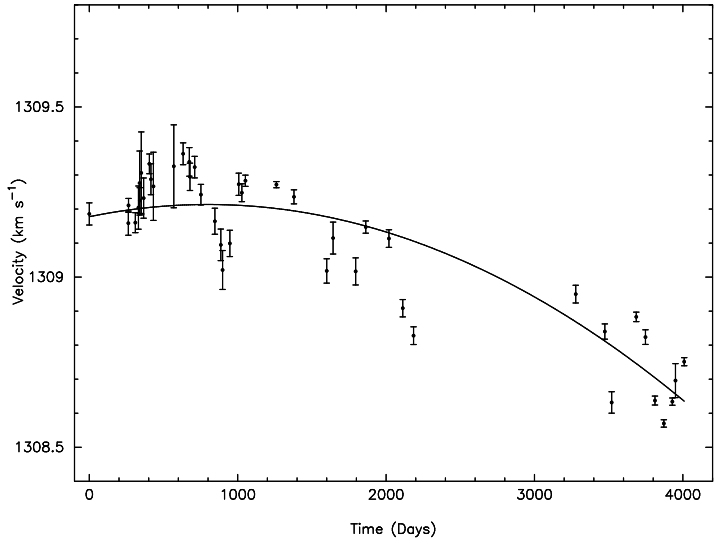}
\caption{ The velocity drift of the ``1309'' km~s$^{-1}$ feature versus time.
The curve has the shape of the 1830 year period expected for this feature. Updated from Humphreys et al. (2004).}
\end{center}
\end{figure}

Over a long period of time the accelerations of the maser features should show deviations from a constant value. That is, the line of sight velocity should be given by a sine wave of about 1,830 years, that is,

\begin{equation}
 v=\sqrt{\frac{GM}{r}} \cos [\Omega_r (t - t_o)]~,
\end{equation}

\noindent
where $\Omega_r$ is the orbital frequency at radius $r$, and $t_o$ is the time when the maser crosses
the midline. The feature at ``1309'' km~s$^{-1}$ has been particularly long lasting, and we have tracked its velocity over a period of more than 12 years now (see Figure 4). The expected velocity function for a central mass of 3.8 $\times$ $10^{6}M_{\odot}$ is shown. There is not enough of a time line to prove convincingly that the curvature is real. The confirmation of sinusoidal variation would be another powerful confirmation of the robustness of the simple dynamical model. 

\section{The disk thickness}\label{sec:diskthick}

The first VLBI data suggested that the disk thickness was less than about 10 $\mu$as (Moran et al., 1995). The statistical distribution of the maser features from the midline can be most reliably estimated by examining the features at azimuth angles near the center of the bowl where changes in radial distance do not cause vertical projected displacements. In Figure 5 we show a histogram of these projected vertical displacements. The distribution is nearly a Gaussian function with a width (FWHM) of 12 microarcseconds. If this represents the real thickness of a disk that is in hydrostatic equilibrium, then the sound speed is about 1.5 km~s$^{-1}$ and the temperature is about 600K (Argon et al., 2007). Note that this may be a slight overestimate since there may be some effects of projection and measurement noise included. This temperature is quite appropriate for the excitation of water maser emission. The temperature could be lower if part of the support originates from magnetic pressure. For magnetic pressure to completely support this disk would require a magnetic field of  about 100 mG, which is about the current upper limit on its measurement (Modjaz et al., 2005).  Note,  however,  that it has been suggested that the masers form in a skin on a thicker disk. It might be possible to test this model if, say, the redshifted masers are generated on the top side of the disk and the blue on the bottom. 
\begin{figure}[!hb]
\includegraphics[height=4.9in,width=1.12in,angle=-90]{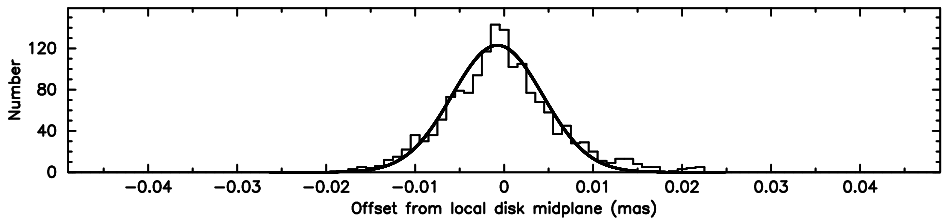}
\vspace{10pt}
  \caption{ A histogram of the vertical positions of the maser features in the systemic group stronger than 250 mJy between velocities of 485 and 510 km~s$^{-1}$, corresponding to azimuth angles of 2 to 6 degrees, where the 
masers originate near the bottom of the bowl.  The average rms measurement error in the relative positions is 
about 1 $\mu$as, which is negligible for this analysis.  From Argon et al. (2007).}
\end{figure}

\section{Prospects for distance measurements}

For a simple flat disk, where the high velocity features lie on the midline and the systemic features are at a fixed radius, the distance can be estimated by the formula

\begin{equation}
 D = \frac{v_H^2}{a~b_H~{\rm sin}~i}~,
\end{equation}\vspace{10pt}

\noindent
where $v_H$ and $b_H$ define the point in the PV diagram (see Figure 1) where the line through the systemic features intersects the Keplerian curve of the high velocity features, $a$ is the line-of-sight acceleration of the systemic features, and $i$ is the inclination of the disk. 
 The slope of the systemic features in the position-velocity diagram, $\Omega = dv/db = \sqrt{GM/r^3}$ so $v_H=\Omega b_H$. (See Greenhill 2004 for an alternate formulation). The angular radius of the systemic features can be estimated from $\Omega$ in the case of NGC~4258 where the inclination is nearly 90$^{\circ}$.
  This is the only way to determine the radius of the systemic features when the 
disk is nearly edge-on. Hence, the measurements of $a$ and $\Omega$ provide the critical link between angular and linear scales.  In the case of NGC~4258, the accelerations vary 
from 7.7 to 8.8 km~s$^{-1}$yr$^{-1}$. To first order, an accurate distance can be obtained if the local slope of the PV diagram is used with the corresponding acceleration and inclination. Proper motions of the systemic features determined from VLBI measurents can also be used to estimate the distance independently of the 
accelerations.  However, in the case of NGC~4258, the change in velocity in a year's time is about 5 times the feature linewidth for a systemic feature, whereas the proper motion is only about 30 $\mu$as, or one-tenth of the typical angular resolution. However a clear demonstration of the correlation between the proper motions and accelerations would be an important confirmation of the simple dynamical model. Clearly, to obtain a distance estimate of accuracy of better than about 5 percent for a warped disk that has a range of accelerations among its systemic features and has effects due to spiral structure requires  sophisticated modeling. This work is underway (see Humphreys et al., 2007a).

\end{document}